\documentclass{appolb}
\usepackage{graphicx}
\begin{document}

\title{Universality of random matrix dynamics
\thanks{Presented at the conference 
Random Matrix Theory: Applications in the Information Era, Krakow, 
April 29th - May 3rd 2019}}
\author{Zdzislaw Burda
\address{AGH University of Science and Technology, Faculty of Physics and Applied Computer Science, al. Mickiewicza 30, 30-059 Krak\'ow, Poland}}
\maketitle
\begin{abstract}
We discuss the concept of width-to-spacing ratio which plays the central role
in the description of local spectral statistics of evolution operators in
multiplicative and additive stochastic processes for random matrices. We show that the local spectral properties are highly universal and depend on a single parameter being the width-to-spacing ratio. We discuss duality between the kernel for Dysonian Brownian motion and the kernel for the Lyapunov matrix for the product of Ginibre matrices. 
\end{abstract}

\PACS{02.10.Yn; 02.50.-r; 05.40.-a}

\section{Introduction}  
Local spectral properties of invariant random matrix ensembles are highly universal \cite{m,dg,k,ey}. This means that these properties depend only on the symmetry class of the ensemble or equivalently on the type of invariance of the probability measure. Here we search for an analogous principle for stochastic processes in the matrix space. We consider prototypes of additive and multiplicative stochastic processes in the space of Hermitian matrices. We show that local spectral statistics of evolution operators for these processes is described by a determinantal point process with the kernel that interpolates between the picket-fence kernel and the sine kernel in a universal way that is controlled by a single parameter being the width-to-spacing ratio \cite{abk1,abk2,abk3}.

The paper is organised as follows. In Section \ref{sec:DBM} we recall Dyson Brownian motion \cite{d1}. In Section \ref{sec:DBMlocal} we evoke an analytic formula for the kernel of Dyson Brownian motion with the initial condition given by equidistant eigenvalues \cite{j}. This result is used in comparative studies towards the end of the paper. In Section \ref{sec:M} we introduce a multiplicative stochastic evolution in the matrix space. In Section \ref{sec:ML} we investigate local statistics of the Lyapunov spectrum associated with this evolution. In Section \ref{sec:Duality} we discuss duality and universality of the kernels of evolution operators for additive and multiplicative stochastic processes. The material presented in Sections \ref{sec:M},\ref{sec:ML},\ref{sec:Duality} is based on a joint work with Gernot Akemann and Mario Kieburg \cite{abk1,abk2,abk3}. The paper is concluded in Section \ref{sec:Conclusions}.
 
\section{Additive matrix evolution - Dysonian random walk \label{sec:DBM}}

We first recall the Dyson construction of additive random walk in the space of matrices \cite{d1}. Let $X_m$ be $N\times N$ complex matrices. The random walk $X_0 \rightarrow X_1 \rightarrow \ldots \rightarrow X_M$ is defined by the recursive formula 
\begin{equation}
X_m = X_{m-1} + \sigma G_m \ .
\label{additive}
\end{equation}
which describes incremental random changes of matrices $X_m$ at discrete times
$m=1,2,\ldots, M$. The increments $G_m$'s are independent identically distributed $N \times N$ Ginibre matrices whose entries are themselves independent identically distributed standard complex variables ${\mathcal C} \mathcal{N}(0,1)$ \cite{g}. $\sigma$ is a scale parameter. One is interested in the evolution of eigenvalues of the Hermitian matrix
$A_m$ associated with $X_m$, which is obtained by the Hermitian projection $A_m = (X_m + X_m^\dagger)/\sqrt{2}$. The evolution equation for this matrix
\begin{equation}
A_m = A_{m-1} + \sigma H_m \ .
\label{AM}
\end{equation}
is analogous to Eq. (\ref{additive}), except that the increments $H_m=(G_m + G_m^\dagger)/\sqrt{2}$ are GUE matrices in this case. The matrix $A_M$ at time $M$ is a sum of the initial matrix and of i.i.d. Gaussian increments
\begin{equation}
A_M = A_0 + \sigma (H_1+H_2+\ldots+H_M) \ .
\label{A0_sum}
\end{equation}
The matrix $A_M$ has $N$ real eigenvalues $a_{Mj}$, $j=1,\ldots, N$. The process of evolution of these eigenvalues is known as Dysonian random walk. One can define physical time $t = M \Delta t$ where $\Delta t$ is the time interval between two consecutive instances of the discrete process. If the scale parameter scales as 
$\sigma = \sigma_c \sqrt{\Delta t}$ where $\sigma_c$ is a positive constant
one can take the limit $\Delta t \rightarrow 0$ to obtain a continuous 
Dyson random walk which is commonly known as Dyson Brownian motion. 
It follows from the stability of GUE matrices \cite{v,bp} that the sum
of i.i.d. increments in Eq. (\ref{A0_sum}) has for $N\rightarrow \infty$ the same limiting eigenvalue density\footnote{One has to divide out a trivial scaling factor $\sqrt{N}$ which is proportional to the width of the eigenvalue distribution. The matrix $H/\sqrt{N}$ has a limiting density.} as a single GUE matrix $\sqrt{M} H$ with the scale parameter $\sqrt{M}$. Eigenvalues of $A_M$ at time $M$ have the same distribution as eigenvalues of the matrix 
\begin{equation}
\tilde{A}_M = A_0 + \sigma \sqrt{M} H = A_0 + \sigma_c \sqrt{t} H\ .
\label{free_sum}
\end{equation}

\section{Local spectral properties of Dyson Brownian motion \label{sec:DBMlocal}}

Using the Dyson Coulomb gas representation \cite{m} of Eq.(\ref{AM}) one can derive the following equations for eigenvalues \cite{d1}
\begin{equation}
a_{m,j} - a_{m-1,j} =  \sum_{k\ne j}\frac{1}{a_{m-1,j}-a_{m-1,k}} + \sigma g_{m,j}
\label{discrete}
\end{equation}
where $g_{m,j}$, $m=1,2,\ldots,M$, $j=1,2,\ldots,N$, is a set of independent standard real normal random variables ${\mathcal N}(0,1)$. The corresponding equations
in the continuous time formalism read
\begin{equation}
d a_j(t) = \sum_{k\ne j}\frac{1}{a_j(t)-a_k(t)} dt  + \sigma_{c} dW_j(t)
\label{coulomb_gas} 
\end{equation}
where $W_j(t)$, $j=1,\ldots,N$ are independent Wiener processes. If one interpretes eigenvalues $a_j(t)$, $j=1,\ldots,N$, as positions of $N$ particles in one dimension at time $t$, then the equations (\ref{coulomb_gas}) describe the Brownian motion of these particles which interact with each other. The potential of the interactions is logarithmic $\ln |a_j - a_i|$. One calls the system "Coulomb gas" since the logarithmic potential is the Coulomb potential in two dimensions. Even if this is a slight abuse of terminology, as the system in question is one-dimensional, the term "Coulomb gas" perfectly reflects the behaviour of the system which imitates thermal behaviour of a gas of repelling particles. Particles' trajectories generated by Eq. (\ref{coulomb_gas}) are continuous. The repulsion potential $\ln|a_j - a_i|$ prevents the trajectories from intersecting each other so if $a_1(t)< a_2(t)< \ldots < a_N(t)$ at some $t$ then $a_1(t')< a_2(t')< \ldots < a_N(t')$ at any later time $t'>t$.  

To solve the differential stochastic equation (\ref{coulomb_gas}) means to determine 
the probability density function, $P_N(x_1,x_2,\ldots, x_N;t)$ which is directly related to the probability $P_N(x_1,x_2,\ldots, x_N;t) dx_1\ldots dx_N$ of finding eigenvalues $a_1, a_2, \ldots, a_N$ at time $t$ in the infinitesimal neighbourhood of $x_1$, $x_2$, $\ldots$, $x_N$. The standard way of solving the problem is to write down the Fokker-Planck equation associated with the stochastic differential equations (\ref{coulomb_gas}) and to solve it for $P_N$. One can then calculate correlation functions \cite{gmw}
\begin{equation}
R_{k}(x_1,x_2,\ldots,x_k;t) = \frac{N!}{(N-k)!} 
\int \ldots \int dx_{k+1} \ldots dx_N P_N(x_1,x_2,\ldots,x_N;t)
\label{Rk}
\end{equation}
which are just appropriately normalised marginal distributions of $P_N$. They can be interpreted as probability densities that $k$ eigenvalues lie in the infinitesimal neighbourhood of $x_1,\ldots,x_k$, except that the total integral of $R_k$ is not one but $N!/(N-k)!$. In particular the first correlation function $R_1(x)$ gives the distribution
of eigenvalues normalised to the number of eigenvalues $\int R_1(x) dx = N$. 

Generally it is difficult to find a closed-form solution to the stochastic differential equations (\ref{coulomb_gas}) since the evolution of the system is very complex and non-stationary. The repulsion makes the gas continuously expand. Details of this expansion are sensitive to the initial positions of particles and the statistical noise. An exception is the situation when the gas is uniformly distributed on the whole real axis (for $N=\infty$) since in this case the effect of expansion is eliminated and the average distance between particles stays on average constant over time. An explicit solution can be found in this case \cite{j}. 

\begin{figure}
\centerline{
\includegraphics[width=6.0cm]{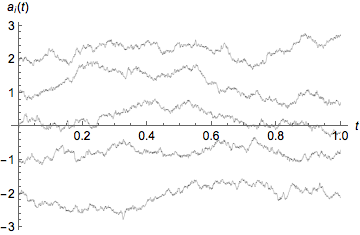} 
\includegraphics[width=6.0cm]{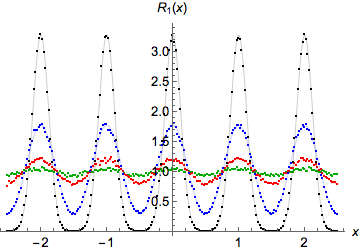}}
\caption{Dyson Brownian motion of eigenvalues of Hermitian matrix which for $t=0$ is diagonal and has equidistant eigenvalues $\lambda_j = j - K$ for $j=1,\ldots,2K-1$, where $N=2K-1$. The eigenvalue spacing is $s=1$ initially. We set $\sigma_c=1$ so the width-to-spacing ratio is $w=\sqrt{t}$ (\ref{wsr_rw}). In the left panel we plot a single realisation of the stochastic evolution (\ref{discrete}) of five central eigenvalues of the matrix which initially, for $t=0$, are located at $\{-2,-1,0,1,2\}$. The matrix size is $N=255$. The right panel shows the central part of the spectral density for $x\in [-2.5,2,5]$ for $w=\sqrt{t}=\{0.125,0.25,0.5,1.0\}$. Solid lines represent the limiting density for $N\rightarrow \infty$ calculated from the analytic formula $R_1(x)=K_t(x,x)$ (\ref{kernel_t}). Different colors correspond to different values of the width-to-spacing ratio parameter (\ref{wsr_rw}): $w=0.125$ (Black), $w=0.25$ (Blue), $w=0.5$ (Red) and $w=1.0$ (Green). Points represent results of Monte-Carlo simulations for $N=255$. For each $w$ we generated $10^5$ matrices.}
\label{Fig:random_walk}
\end{figure}

This situation can be imitated by a finite-$N$ system with the initial condition $a_j(0) = (j - K) s$, $j=1,2,\ldots, N$ with $N=2K-1$, which describes $N$ equidistant eigenvalues
(particles) uniformly distributed on the real axis within the boundaries $-s(K-1)$ and $s(K-1)$. This can be realised by choosing a diagonal matrix $A_0 = \mbox{diag}\left(-s(K-1),\ldots, -s,0,s,\ldots, s(K-1)\right)$ in Eq. (\ref{A0_sum}).
During the evolution (\ref{coulomb_gas}) eigenvalues drift away from each other. The peripheral eigenvalues move away the fastest. The further an eigenvalue is from the gas boundary the slower it moves since it is confined by eigenvalues on both sides which have to drift away first. When $N$ is large the mean spacing between internal eigenvalues is almost constant and equal to the initial spacing $s$ for a long time $t$, or more precisely for time $t \ll N s^2/\sigma_c^2$. This can be seen from the following argument. The width of the eigenvalue distribution (radius of gyration) is equal to the square root of the second cumulant of the eigenvalue distribution of the matrix $\tilde{A}_t=A_0 + \sigma_c \sqrt{t} H$ (\ref{free_sum}). For large $N$ the second cumulant of the eigenvalue distribution of $\tilde{A}_t$ can be approximated as a sum of the second cumulant of $A_0$ which is $(sN)^2/12$ and of $\sigma_c \sqrt{t} H$ which is $\sigma_{c}^2 N t$, since for large $N$ the addition of these matrices is almost free \cite{v}. This gives $(sN)^2/12 + \sigma_c^2 N t$. Let $S$ be the spacing of a hypothetical distribution of $N$ equidistant particles with the same radius of gyration $(S N)^2/12 = (sN)^2/12 + \sigma_{c}^2 N t$. This hypothetical spacing is related to the initial spacing $s$ as $S = s\sqrt{1 + 12 \sigma_c^2 t/(s^2 N)}$. Clearly $S$ gives the upper bound on the spacing between eigenvalues of the matrix (\ref{free_sum}) in the center of the spectrum. The initial spacing $s$ gives the lower bound (since eigenvalues repel). For fixed $t$, the upper bound $S$ approaches $s$ for $N\rightarrow \infty$. This means that the mean spacing between eigenvalues in the center of the spectrum is equal $s$ in this limit. The same holds also in the double scaling limit: $t=t(N)$ and $N\rightarrow \infty$, as long as $t=t(N)$ grows slower than $N$, that is $t\sim o(N)$. More generally for $N\rightarrow \infty$ one can assume that the spacing between eigenvalues lying in any compact interval is constant and equal $s$. This is an enormous simplification. In effect one can give an explicit closed-form solution of the evolution equations (\ref{coulomb_gas}) for eigenvalues in the bulk in the limit $N\rightarrow\infty$. The solution was given in \cite{j} where it was shown that the correlation functions (\ref{Rk}) have the determinantal form 
\begin{equation}
R_k(x_1,\ldots,x_k; t) = \det K_t(x_i,x_j)_{i,j=1,\ldots,k}
\end{equation}
with the kernel
\begin{equation}
K_t(x,y) = \frac{1}{\pi s} \mbox{Re} \sum_{k=-\infty}^{\infty}
\exp\left[-2 \pi^2 w^2  k(k-1) \right] \frac{\exp[i\pi \left((2k-1) x/s + y/s\right)]}{2 \pi w^2  k + i (y-x)/s} \ .
\label{kernel_t}
\end{equation}
The corresponding eigenvalue distribution is $R_1(x)=K_t(x,x)$.
The evolution of the eigenvalue distribution with time is shown in Fig.\ref{Fig:random_walk} where we plot the limiting density for $N=\infty$ derived analytically $R_1(x) = K_t(x,x)$ from Eq. (\ref{kernel_t}) and the corresponding histograms for $N=255$ obtained by Monte-Carlo simulations of Eq. (\ref{free_sum}).
One can see that the histograms for $N=255$ coincide with the limiting density. This means that the mean spacing between these five eigenvalues remains almost constant for the given evolution times $t$, in agreement with the argument given above.
The kernel $K_t$ (\ref{kernel_t}) depends on time $t$ through the parameter
\begin{equation}
w = \frac{\sigma_{c} \sqrt{t}}{s} \ .
\label{wsr_rw}
\end{equation}
This parameter has a clear physical meaning. The numerator $\sigma_c\sqrt{t}$ is approximately equal to the width of the peak representing the probability of finding an eigenvalue that undergoes Brownian motion between neighbouring eigenvalues, while the denominator $s$ is equal to the average spacing between eigenvalues. For this reason we call $w$ width-to-spacing ratio. For short times the evolution of individual eigenvalues is described by an almost free Brownian motion and the peaks are Gaussian. When the peaks get broader the repulsion starts to deform them. The kernel (\ref{kernel_t}) depends on the positions $x$ and $y$ through the combinations $x/s$ and $y/s$. One can express $x$ and $y$ in  units of $s$. This amounts to introducing rescaled variables $\xi=x/s$ and $\zeta=y/s$. Denote the resulting kernel by $K_w(\xi,\zeta)$. It is related to the kernel (\ref{kernel_t}) as $K_w(\xi,\zeta) = s K_t(s\xi,s\zeta)$. The prefactor $s$ is equal to the Jacobian $dx/d\xi$. The kernel $K_w$ is 
\begin{equation}
K_w(\xi,\zeta) = \frac{1}{\pi} \mbox{Re} \sum_{k=-\infty}^{\infty}
\exp\left[-2 \pi^2 w^2  k(k-1) \right] \frac{\exp[i\pi \left((2k-1) \xi + \zeta\right)]}{2 \pi w^2  k + i (\zeta-\xi)} \ .
\label{Kw1}
\end{equation}
The width-to-spacing ratio $w$ increases as time goes on. The peaks of the distribution are initially localised at integers but they broden when $w$ increases. They begin to overlap when $w$ is of order one. When $w$ further increases the gap between peaks closes and the spectrum flattens  (see Fig.\ref{Fig:random_walk}). Eventually in the limit $w \rightarrow \infty$ the spectrum becomes flat. The density $R_{1,w}(\xi) = K_w(\xi,\xi)$
(\ref{kernel_t}) interpolates between the Dirac-delta picket fence
\begin{equation}
R_{1,w=0}(\xi) = \sum_{j=-\infty}^{\infty}\delta(\xi-j)
\label{R1dirac}
\end{equation} 
for $w=0$, and a fully translationally invariant flat distribution  
\begin{equation}
R_{1,w=\infty}(\xi) = 1 
\label{R1flat}
\end{equation} 
for $w\rightarrow \infty$. The limiting kernel is just the standard sine kernel \cite{m}
\begin{equation}
K_{w=\infty}(\xi,\zeta)= \frac{\sin\left(\pi(\xi-\zeta)\right)}{\pi(\xi-\zeta)} \ .
\label{Ksine}
\end{equation}
in this case.

\section{Multiplicative matrix evolution \label{sec:M}}

\begin{figure}
\centerline{
\includegraphics[width=6.0cm]{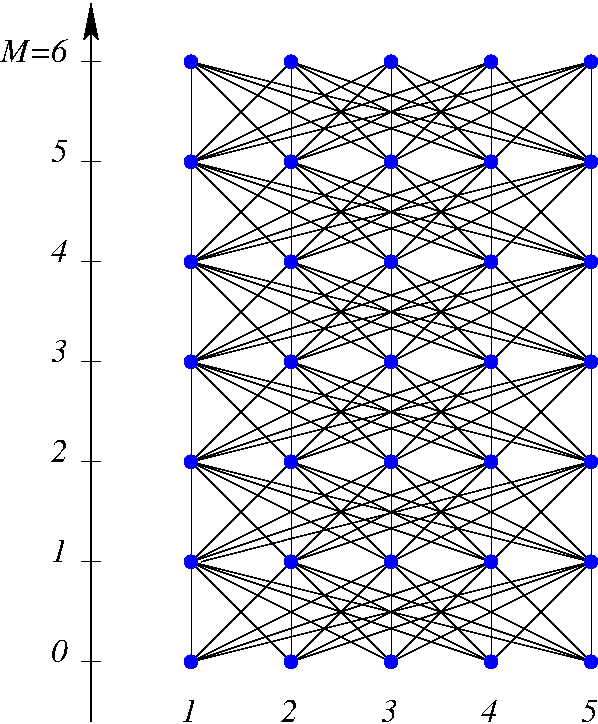}} 
\caption{Schematic representation of the architecture of a multilayered system. 
Nodes (blue dots) in a layer $m$ represent components of the state vector $| x \rangle_m$ of the system at time $m$. The state $|x \rangle_m$ is obtained from $|x \rangle_{m-1}$ by a linear map $|x \rangle_m = G_m| x \rangle_{m-1}$. Elements $(G_m)_{ij}$ of the transfer matrix $G_m$ are represented by edges of the network. The network shown in the figure represents signal processing of $N=5$ degrees of freedom in $M=6$ time steps.}
\label{Fig:mls}
\end{figure}

Let us now consider a multiplicative stochastic matrix evolution defined by the following recursive formula 
\begin{equation}
X_m = G_m X_{m-1} 
\label{multiplicative}
\end{equation}
where as before $m=1,\ldots,M$ is a discrete time index and the random increments
$G_m$'s are independent identically distributed $N \times N$ Ginibre matrices. This equation is analogous to Eq. (\ref{additive}) except that the incremental changes are now multiplicative. One can analytically determine the eigenvalue distribution of $X_M$ \cite{bjw,ab}. Here we are interested in the Hermitian matrix $Y_M=X_M^\dagger X_M$ associated with $X_M$. 
For the multiplicative process (\ref{multiplicative}) $Y_M$ is a more natural Hermitian partner of $X_M$ than $(X_M^\dagger + X_M)/\sqrt{2}$ that was used for the additive process (\ref{additive}). Clearly eigenvalues of $Y_M$ correspond to squares of singular values of $X_M$. Let us for simplicity assume that $X_0$ is an identity matrix. In this case $Y_M$ is 
\begin{equation}
Y = (G_M G_{M-1}\ldots G_1)^\dagger (G_M G_{M-1}\ldots G_1) \ .
\label{YM}
\end{equation}
The eigenvalue distribution of this matrix was determined in \cite{akw}.
From here on we skip the index $M$ and for brevity write $Y$, to simplify notation. We are interested in the evolution of eigenvalues $y_{Mj}$, $j=1,\ldots,N$, of the matrix $Y$ or alternatively in the evolution of Lyapunov exponents $\lambda_{Mj}$, $j=1,\ldots,N$, that is eigenvalues of the Lyaponov matrix \cite{abk1,abk2,abk3}
\begin{equation}
L = \frac{1}{2M} \log (G_M G_{M-1} \ldots  G_1)^\dagger (G_M G_{M-1} \ldots G_1) = \frac{1}{2M}
\log Y \ .
\label{lyapunov}
\end{equation}
For any finite $M$ and $N$ the spectra of $L$ and $Y$ contain exactly the same information since $y_{Mj} = e^{2M \lambda_{Mj}}$. The product $G_M G_{M-1}\ldots G_1$ can be viewed as a discrete time evolution operator or a transfer matrix in a system with $N$ degrees of freedom. An initial state of the system $| x \rangle_0$ is mapped onto the state  
\begin{equation}
\left| x \right\rangle_M = G_M G_{M-1}\ldots G_1 \left|x\right\rangle_0
\end{equation}
at time $M$. This equation can be depicted symbolically as a multilayered network as sketched in Fig.\ref{Fig:mls}.
The layout of this network is typical for signal processing in artificial neural networks known from machine learning. Here the signal processing from layer to layer 
$\left| x \right\rangle_m = G_m \left| x \right\rangle_{m-1}$ is linear while in neural networks it is non-linear. As we shall see even for the linear case the system undergoes an interesting phase transition between "deep" systems and "shallow" ones which manifests as a change of local spectral statistics of Lyapunov exponents in the limit $M,N\rightarrow \infty$.

Let $M=M(N)$ be a monotonically increasing function of $N$ and let $a$ be the limiting aspect ratio of the system
\begin{equation}
a = \lim_{N\rightarrow \infty} a_N = \lim_{N\rightarrow \infty} \frac{N}{M(N)} \ .
\label{aspect_ratio}
\end{equation}
Depending on the value of $a$ one can distinguish three types of architecture: deep systems for $a=0$, shallow systems for $a=\infty$ and critical ones for $0<a<\infty$.
When the number of time slices $M$ super-linearly grows with the number of degrees of freedom, $N$, {\em e.g.} $M\sim N^2$, the limiting system is deep; when it scales sub-linearly, {\em e.g.} $M\sim \sqrt{N}$, the limiting system is shallow. The architecture is critical when $M$ is proportional to $N$. For large but finite $M,N$ the system can be called deep when $M \gg N$ and shallow when $M \ll N$.

\section{Local statistics of Lyapunov spectrum \label{sec:ML}}

Eigenvalues of the Lyapunov matrix (\ref{lyapunov}) for the product of GUE matrices assume deterministic values \cite{n1,ni}
\begin{equation} 
\lambda_j = \frac{\psi(j)}2, \quad j=1,\ldots, N 
\label{positions}
\end{equation}
in the limit $M\rightarrow \infty$ where $\psi(z)= \left(\log \Gamma(z)\right)'$ is the digamma function. For finite $M$ but very large $M\gg N$ eigenvalues of the Lyapunov matrix (\ref{lyapunov}) have a probability distribution that can be approximated by a sum of Gaussian peaks centered around the limiting values \cite{abk1,abk2}
\begin{equation}
R_{1}(\lambda) \approx \sum_{j=1}^N \frac{1}{\sqrt{2\pi \sigma_j^2}} \exp\left[-\frac{(\lambda-\lambda_j)^2}{2\sigma_j^2}\right] \ .
\label{rhoMN}
\end{equation}
Each peak is normalised to one, so the total distribution is normalised to the number of eigenvalues $N$. The widths of the peaks depend on the derivative of the digamma function
\begin{equation} 
\sigma_j = \frac{\sqrt{\psi'(j)}}{4M}, \quad j=1,\ldots, N \ .
\label{widths}
\end{equation}
For $M\rightarrow \infty$ the peaks become Dirac deltas. The distribution (\ref{rhoMN}) has an interesting property. The positions and widths of the peaks do not depend on $N$. This means that when $N$ is increased new peaks are added to the distribution but the old ones stay intact. 

The digamma function satisfies the following identity $\psi(z+1) = \psi(z) + 1/z$. 
Thus the mean spacing between neighbouring Lyapunov exponents is
\begin{equation}
\lambda_{j+1}-\lambda_j = \frac{1}{2j} \ .
\end{equation}
The digamma function has the asymptotic expansion of $\psi(z) = \ln z + 1/(2z) -1/(12z^2) + \ldots$ for $\mbox{Re}(z)>0$. In consequence, the width of the $j$-th peak is $j \sim \sqrt{\psi'(j)/(4M)} \approx \sqrt{1/(4jM)}$. This means that for large $j$ the width-to-spacing ratio can be approximated by 
\begin{equation}
w_{j} = \frac{\sigma_{j+1}+\sigma_j}{2(\lambda_{j+1}-\lambda_j)} \approx \sqrt{\frac{j}{M}} \ .
\label{wsrj}
\end{equation}
The width-to-spacing ratio increases when $j$ increases. It is maximal for $j$ at the upper end of the spectrum, where it takes the value $\sqrt{N/M}=\sqrt{a_N}$ (\ref{aspect_ratio}). 

We are now going to discuss local spectral statistics of the Lyapunov exponents in the limit $M,N \rightarrow \infty$. We start from an explicit expression for the kernel 
of the matrix $Y$ (\ref{YM}) for finite $M$ and $N$ \cite{abk2,abk3}
\begin{equation}
K_Y(x,y) = \frac{1}x\, \sum_{j=1}^{N} \left(\frac{x}y\right)^j G_j(y),
\label{eq:main1}
\end{equation}
where
\begin{equation}
G_j(y) =\int_{-i\infty}^{+i\infty} \frac{dt}{2\pi i}\frac{\sin (\pi t)}{\pi t}\ {y}^t 
\left(\frac{\Gamma(j-t)}{\Gamma(j)}\right)^{M+1} \frac{\Gamma(N-j+1+t)}{\Gamma(N-j+1)}.
\label{eq:main2}
\end{equation}
There are many equivalent expressions for the kernel that can be found in the literature on the subject \cite{akw,aik,kz,lwz}. The one given above has been derived from a formula 
in \cite{aik}. An advantage of the integral representation (\ref{eq:main2}) is that it is very well suited for taking various limits $M\rightarrow \infty$ and  $N\rightarrow \infty$. One can for example easily transform the kernel $K_Y$  (\ref{eq:main1},\ref{eq:main2}) to the kernel $K_L$ for the matrix $L$ (\ref{lyapunov}) by changing variables in (\ref{eq:main1}). By doing this one can immediately recover Eq.(\ref{rhoMN}) from the asymptotic behaviour of the integrand (\ref{eq:main2}) for $M\rightarrow \infty$ \cite{abk2,abk3}. Here we are mainly interested in the double scaling limit $N,M\rightarrow \infty$ and $N/M \rightarrow a$, for $a$ (\ref{aspect_ratio}) being a finite and positive number, $0<a<\infty$, which corresponds to the critical scaling. 
 
The number of Lyapunov exponents between $x$ and $x+dx$ is proportional to the eigenvalue density $\rho_\lambda(x)dx$. The mean spacing between Lyapunov exponents in the neighbourhood of $x$ is inversely proportional to $\rho_\lambda(x)$ so it depends on the position $x$ in the spectrum. It is convenient to make the spacing independent of the position in the spectrum. One does it by unfolding the spectrum, {\em i.e.} by expressing the distribution in the variable 
\begin{equation}
p=\int^\lambda_{-\infty} \rho_\lambda(x) dx
\label{cdf}
\end{equation}
which has the uniform distribution on the unit interval \cite{gmw}. For finite $N$ this variable can be imitated by $p = j/N$ where $j$ is the index of the Lyapunov exponent $\lambda_j$. Since Lyapunov exponents are ordered $\lambda_1< \lambda_2 < \ldots <\lambda_N$, the quantity $p=j/N$ can be interpreted as the probability of finding an exponent smaller than or equal $\lambda_j$: $\mbox{Prob}(\lambda\le \lambda_{j}) = j/N = p$. For  $N\rightarrow \infty$ the last equation takes the form (\ref{cdf}), which means that the variable $p=j/N$ indeed unfolds the spectrum in the limit $N\rightarrow \infty$. The eigenvalue density $\rho_\lambda(x)dx$ is known analytically for any finite $M$ \cite{n2} but unfortunately it is expressed in an intricate parametric form from which it is hard to reconstruct the unfolding map. However for $M \rightarrow \infty$ one can find another way to unfold the spectrum \cite{n1}. It is based on the asymptotic behaviour of Lyapunov exponents $\lambda_j = \log(j)/2 +o(1/j)$ for large $j$ that we discussed above. A consequence of this asymptotic behaviour is that the quantity $u_j = e^{2\lambda_j}/N$ behaves asymptotically as $u_j = j/N (1+ o(1/j)) \approx p$. Thus, for $j$ of order $N$ it unfolds the spectrum when $N\rightarrow \infty$. The variables $u_j$ can be viewed as eigenvalues of the matrix 
\begin{equation}
u = \frac{e^{2L}}N = \frac{Y^{1/M}}N \ .
\label{uLY}
\end{equation}
For $M,N\rightarrow \infty$ the eigenvalue spectrum of $u$ becomes uniform on $(0,1)$ and thus it unfolds the Lyapunov spectrum. 

The kernel $K_u(p_x,p_y)$ for the unfolded spectrum can by obtained from $K_Y(x,y)$ (\ref{eq:main1}) by changing variables to $p_x=x^{1/M}/N$, $p_y=y^{1/M}/N$ as follows from (\ref{uLY}). This amounts to replacing $x$ and $y$ by $x = (p_x N)^M$ and $y=(p_y N)^M$ 
in $K_Y(x,y)$. One has also to include the Jacobian $dx/dp_x$ in the transformation law
$K_u(p_x,p_y) = dx/dp_x K_Y(x,y)$. The mean spacing between eigenvalues of the uniform spectrum on the unit interval is $1/N$, so if one wants to investigate local level statistics at a point $p$ of the unfolded spectrum one has to zoom in at this point to the local scale 
\begin{equation}
p_x = p + \frac{\xi}{N} \ , \quad p_y= p+\frac{\zeta}{N}
\end{equation}
where $\xi$ and $\zeta$ are of order one. One can now take the double scaling limit $N\rightarrow \infty$, $N/M(N)\rightarrow a$ keeping the aspect ratio (\ref{aspect_ratio}) finite and positive $0<a<\infty$. We denote the limiting kernel for the unfolded spectrum at the point $p \in (0,1)$ by
\begin{equation}
K_p(\xi,\zeta) = \lim_{N\rightarrow \infty, N/M\rightarrow a} 
K_u\left(p + \frac{\xi}{N}, p+\frac{\zeta}{N}\right) \ . 
\end{equation}   
The result reads \cite{abk2}
\begin{equation}
K_{p}(\xi,\zeta) = \frac{1}{2\pi ap} \mbox{Re} \sum_{\nu=-\infty}^{+\infty}
\exp\left(\frac{\nu(\xi-\zeta)}{ap}\right) \mbox{erfi} \left(\frac{\pi\sqrt{2ap}}2 + i\frac{\zeta-\nu}{\sqrt{2ap}} \right) \ .
\label{Kp}
\end{equation}
where $\mbox{erfi}$ is the imaginary error function. The details of the calculations are presented in \cite{abk3}. Here we only give a  short recap. One begins with an explicit expression for the kernel $K_Y$ of the matrix $Y$ for finite $M$ and $N$, as for instance the one given here by Eqs. (\ref{eq:main1}) and (\ref{eq:main2}). By changing variables $Y \rightarrow u$ (\ref{uLY}) one can then determine the kernel $K_u$ of the $u$-spectrum which becomes unfolded in the limit $M,N\rightarrow \infty$. Before one takes the limit one has to zoom in at a point $p$ of the $u$-spectrum. Eventually one takes the double scaling limit $N\rightarrow \infty$ and $a_N=N/M \rightarrow a$, which can be done by replacing $M$ by $N/a$ and then taking the limit $N\rightarrow \infty$. 

The resulting expression (\ref{Kp}) depends on the product $ap$ of the aspect ratio $a$ and the position in the spectrum $p\in (0,1)$. The combination $\sqrt{ap}$ can be easily identified from Eq. (\ref{wsrj})
\begin{equation}
w_{j=pN} = \sqrt{\frac{j}{M}} = \sqrt{a_Np} \rightarrow \sqrt{ap}
\end{equation}
as the width-to-spacing ratio at the position $p$ of the spectrum. For brevity we denote it by $w=\sqrt{ap}$. The kernel (\ref{Kp}) for the given width-to-spacing ratio is
\begin{equation}
\hat{K}_{w}(\xi,\zeta) = \frac{1}{2\pi w^2} \mbox{Re} \sum_{\nu=-\infty}^{+\infty}
\exp\left(\frac{\nu(\xi-\zeta)}{w^2}\right) \mbox{erfi} 
\left(\frac{\pi w}{\sqrt{2}} + i\frac{\zeta-\nu}{w \sqrt{2}} \right) \ .
\label{Kw2}
\end{equation}
We denote it here by $\hat{K}_w$ to distinguish it from the kernel $K_w$ (\ref{Kw1}) that was discussed in the previous section. The corresponding eigenvalue density $\hat{R}_{1,w}(\xi) = \hat{K}_{w}(\xi,\xi)$ is
\begin{equation}
\hat{R}_{1,w}(\xi) = \frac{1}{2\pi w^2} \mbox{Re} \sum_{\nu=-\infty}^{+\infty}
\mbox{erfi} \left(\frac{\pi w}{\sqrt{2}} + i\frac{\xi-\nu}{w \sqrt{2}} \right) \ .
\label{bR}
\end{equation}
It interpolates between a picket-fence made of Dirac delta functions 
for $w\rightarrow 0$, and a flat density  for $w\rightarrow \infty$, in the same manner as 
the kernel $K_w$, (\ref{R1dirac}) and (\ref{R1flat}). The limiting form of the kernel $\hat{K}_w$ is given by the sine kernel $w\rightarrow \infty$, the same as for $K_w$ (\ref{Ksine}). Is it a coincidence, or maybe the kernels are equivalent?

\section{Duality and universality \label{sec:Duality}}

It was Jac Verbaarschot and Maurice Duits who first suggested that the two kernels might be identical for any $w$ \cite{vd}. We have checked that this is indeed the case \cite{abk3}. The map between the expressions for $K_w$ (\ref{Kw1}) and for $\hat{K}_w$ (\ref{Kw2}) is provided by the Poisson summation formula which transforms the sum  over $\nu$ in Eq. (\ref{Kw2}) onto the sum over Fourier modes, $k$, in Eq. (\ref{Kw1}). In a sense, the two expressions are dual to each other. The Dirac picket-fence limit (\ref{R1dirac}) corresponds to the large time behaviour of $\hat{K}_w$ and the short time behaviour of $K_w$, while the flat limit (\ref{R1flat}) the other way round. This again reflects the duality of the two kernels.

We have checked by Monte-Carlo simulations \cite{abk2,abk3} that the local spectral density of unfolded Lyapunov spectrum coincides with the limiting density (\ref{bR}) within the numerical accuracy also when one replaces Ginibre matrices $G_m$ in the evolution equation  (\ref{multiplicative}) by random matrices made of i.i.d. non-Gaussian random centered complex variables, or by weakly correlated Ginibre matrices. We refer the interested reader to \cite{abk2,abk3}. This is an indication that the universality of local spectral statistics extends also beyond the realm of Gaussian Markov stochastic processes.  

\section{Conclusions \label{sec:Conclusions}}

We have shown here that the kernels describing local eigenvalue statistics of evolution operators for multiplicative and additive Gaussian stochastic processes in the space of Hermitian matrices (for Dyson index $\beta=2$) interpolate between the Dirac-delta kernel and the sine kernel in a universal way. The interpolation is controlled by the width-to-spacing ratio. It would be interesting to check if a similar universality holds also for real-symmetric ($\beta=1$) and quaternionic matrices ($\beta=4$). Here we concentrated on local spectral statistics in the bulk but one can extend the analysis also to the hard and soft edges of the spectrum \cite{abk2,abk3,lww}.

The main message of the paper is that the spectrum of eigenvalues (or Lyapunov exponents) of the evolution operator constructed as the product of i.i.d. Ginibre matrices changes from a continuous to a discrete one depending on the ratio of the number of degrees of freedom and the propagation time. This spectral change is a prototope of a phase transition that seems to be generic for systems having one distinguished direction for which evolution is driven by the transfer matrix composition rule. Such a situation takes place in many physical systems. Examples include evolution operators in dynamical systems \cite{i}, quantum transport \cite{b}, sequential MIMO systems \cite{m2}, quantum maps \cite{bbtv}, multiplex networks \cite{bnl}, artificial neural networks \cite{s}, thermal field theory \cite{jw}, CDT gravity \cite{ajl} and others. The occurence of such a spectral phase transition seems to be an inherent feature of multilayered systems when they change from shallow to deep ones.

\section*{Acknowledgements}

This contribution is based on a joint work with Gernot Akemann and Mario Kieburg. I like to thank Gernot and Mario for many exciting, illuminating and inspiring discussions which I have enjoyed very much. I also want to thank Jac Verbaarschot and Maurice Duits for drawing our attention to the paper \cite{d1} and for suggesting the equivalence of the kernels.

\end{document}